\documentclass[aps,prl,reprint]{revtex4-1}

\usepackage{amsmath, amssymb, amsfonts}

\usepackage{graphicx}
\usepackage{ae,aecompl}
\usepackage{bm}
\usepackage{textcomp}

\usepackage[colorinlistoftodos, shadow,textsize= footnotesize]{todonotes}

\newcommand{\degree}{\ensuremath{^\circ}}

\begin{document}

\title{Nonlinear atom interferometer surpasses classical precision limit}

\author{C. Gross}
\author{T. Zibold}
\author{E. Nicklas}
\affiliation{Kirchhoff-Institut f\"ur Physik, Universit\"at Heidelberg,
Im Neuenheimer Feld 227, 69120 Heidelberg, Germany}
\author{J. Est\`eve}
\altaffiliation{Present address: Laboratoire Kastler Brossel, CNRS, UPMC, Ecole Normale Sup\'erieure, 24 rue Lhomond, 75231 Paris, France.}
\author{M. K. Oberthaler}
\affiliation{Kirchhoff-Institut f\"ur Physik, Universit\"at Heidelberg,
Im Neuenheimer Feld 227, 69120 Heidelberg, Germany}

\begin{abstract}
Interference is fundamental to wave dynamics and quantum
mechanics. The quantum wave properties of particles are
exploited in metrology using atom interferometers, allowing for
high-precision inertia measurements~\cite{Gustavson:1997aa, Fixler:2007aa}. 
Furthermore, the state-of-the-art time standard is based on an interferometric technique
known as Ramsey spectroscopy. However, the precision of an
interferometer is limited by classical statistics owing to the finite
number of atoms used to deduce the quantity of interest~\cite{Santarelli:1999aa}. 
Here we show experimentally that the classical precision limit can be surpassed
using nonlinear atom interferometry with a Bose-Einstein
condensate. Controlled interactions between the atoms lead to
non-classical entangled states within the interferometer; this
represents an alternative approach to the use of non-classical input
states~\cite{Hald:1999aa,Kuzmich:2000aa,Meyer:2001aa, Leibfried:2004aa, Roos:2006aa}.
Extending quantum interferometry~\cite{Giovannetti:2004aa} to the regime of
large atom number, we find that phase sensitivity is enhanced by
$15$ per cent relative to that in an ideal classical measurement. Our
nonlinear atomic beam splitter follows the ``one-axis-twisting''
scheme~\cite{Kitagawa:1993aa} and implements interaction control using a narrow
Feshbach resonance. We perform noise tomography of the
quantum state within the interferometer and detect coherent spin
squeezing with a squeezing factor of  $\xi_{S}^{2} =  -8.2$dB~\cite{Wineland:1994aa, Sorensen:2001ab, Esteve:2008aa,Schleier-Smith:2010aa,Appel:2009aa}. 
The
results provide information on the many-particle quantum state,
and imply the entanglement of $170$ atoms~\cite{Sorensen:2001aa}.

\end{abstract}

\maketitle

The concept of interferometry relies on the splitting of a quantum
state into two modes, a period of free evolution and, finally, the recombination
of the modes for readout. The key observable, the accumulated
relative phase, is inferred from the measured population
difference between the two output states~\cite{Cronin:2009aa}.  
A prominent example of
linear interferometry is Ramsey spectroscopy, a technique used to
define the current time standard. The concept underlying Ramsey
spectroscopy is shown in Fig. 1a. 
A quantum state of $N$ atoms is
transformed unitarily by a linear beam splitter, resulting in a coherent
spin state  $|\Psi\rangle \propto  (\hat{a}^{\dagger}+ \hat{b}^{\dagger})^{N}|0\rangle$, where $\hat{a}^{\dagger}$ and $\hat{b}^{\dagger}$ are the creation operators for the modes $a$ and $b$.
This beam splitter is typically realized using
a $\pi/2$ microwave pulse coupling two internal atomic states. The mismatch
between the frequency of the microwave field and the frequency
of the atomic transition to bemeasured leads to an accumulated phase, $\varphi$, after the evolution time $\tau$ and to a quantum state $|\Psi\rangle \propto  (\hat{a}^{\dagger}+ e^{i \varphi}\hat{b}^{\dagger})^{N}|0\rangle$. 
To convert the relative phase, $\varphi$, into an observable population difference, $n = (N_{a}-N_{b})/2$, where $N_a$ and $N_{b}$ are the
respective populations of the two modes $a$ and $b$, the two states are coupled again before readout. 
Varying the relative phase by $2\pi$ causes
the observed population difference to change sinusoidally, producing
a so-called Ramsey fringe. The interferometerÕs phase error, $\Delta \varphi = \Delta n/\frac{\partial n}{\partial \varphi}$, is determined by the root mean square error, $\Delta n$, of the population difference and the slope of the interference signal, $\frac{\partial n}{\partial \varphi}$. Best experimental performance is achieved at the zero crossing $(n=0)$
of the Ramsey fringe, where the slope is maximal. \\

In linear interferometry, the phase precision for an $N-$atom coherent spin state can be explained by classical statistics. 
The situation is equivalent to $N$ individual measurements on a single particle. At the most sensitive point of the interferometer, each particle has an equal chance to to be measured in state $|a\rangle$ or $|b\rangle$; the uncertainty for $N$ particles is thus $\Delta n = \sqrt{N}/2$.
The maximal slope $\frac{\partial n}{\partial \varphi} = \mathcal{V} N/2 $ is determined by the visibility $\mathcal{V} $ which is close to one for a macroscopically populated coherent spin state.  
The resulting minimal phase error  for a classical measurement $\Delta \varphi = 1/\sqrt{N}$ is known as the standard quantum limit.
In the case of correlated particles, this classical limit
can be exceeded; doing so is the subject of the emerging field of
quantum metrology. 
The idea is to achieve improved scaling of the
interferometric phase sensitivity with the number of quanta, $N$, which
are typically atoms or photons, using entangled quantum states within
the interferometer~\cite{Giovannetti:2004aa, Pezze:2009aa}. 
Here the fundamental quantum limit is known as Heisenberg limit where the phase error is $\Delta \varphi =1/N$. 

\begin{figure*}[t]
\begin{center}
\includegraphics[width = 170mm]{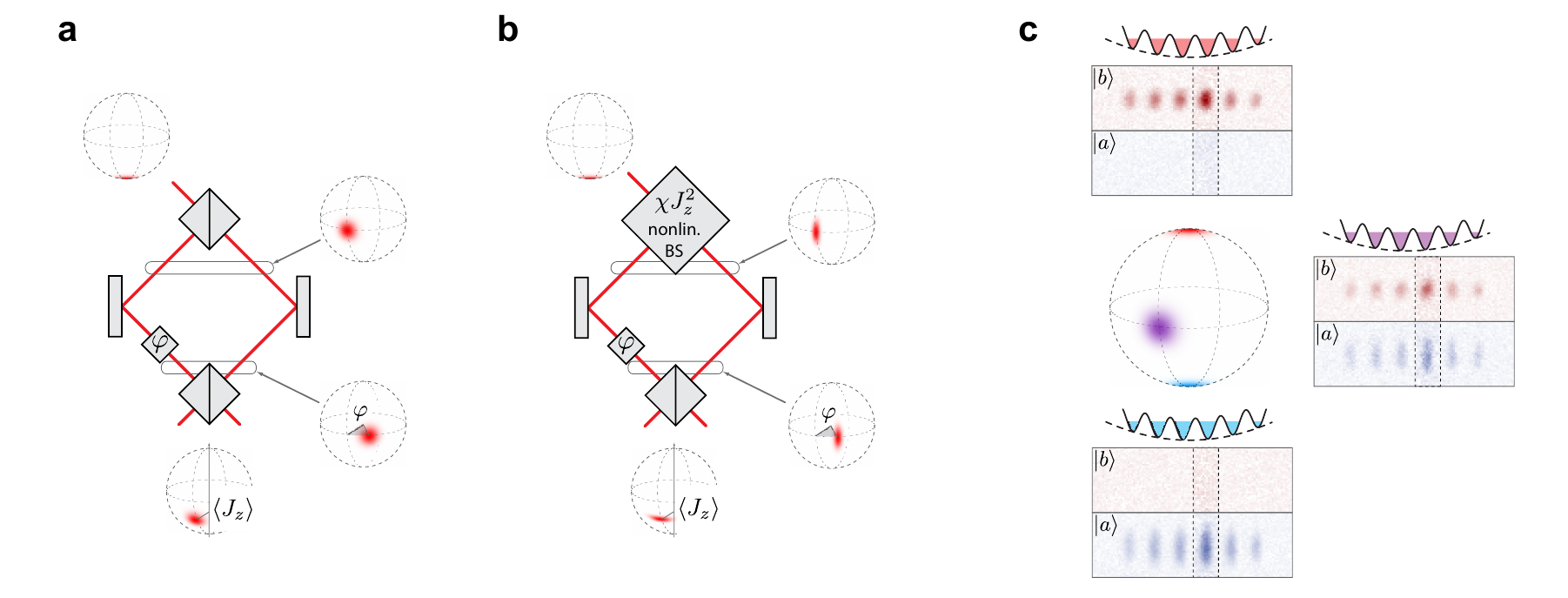}
\caption{
Comparison of linear and nonlinear interferometry. {\bf a, } Schematic of a classical linear interferometer. 
The input state (top) is brought into a superposition of two modes by the first beam splitter which is indicated in the Bloch sphere representation. 
This coherent spin state evolves freely and acquires a phase shift $\varphi$ proportional to the quantity to be measured until
the two modes are mixed by a second beam splitter, where the acquired
relative phase translates into an observable mean population imbalance $\langle J_{z}(\varphi) \rangle$ at the two output ports. 
{\bf b,} Here the input state is split by a nonlinear beam splitter, leading to an entangled quantum state, i.e. a coherent spin squeezed state with reduced fluctuations in the relative phase $\varphi$. 
This phase squeezed state improves the precision of this interferometer beyond the classical limit.
{\bf c,} We prepare six independent Bose-Einstein condensates of $^{87}$Rubidium in a one dimensional optical lattice. 
Two hyperfine states form a two mode system, i.e. the interferometer, in each well. 
The Bloch sphere representation of the quantum state in one well is shown for all atoms in state $|a\rangle$ (blue), in state $|b\rangle$ (red) and in a coherent superposition of the two states (purple). 
An imaging system with high spatial resolution ($\approx 1\,\mu$m) allows the detection of the condensate in each well individually using high intensity absorption imaging~\cite{Reinaudi:2007aa}. 
State-selective time-delayed imaging causes
the atomic clouds to have different shapes. 
}
\label{fig. explanation}
\end{center}
\end{figure*}

In this letter we report on a direct experimental demonstration of interferometric phase precision beyond the standard quantum limit in a novel nonlinear Ramsey interferometer.  
Two hyperfine states labeled $|a\rangle$ and $|b\rangle$ of a $^{87}$Rubidium Bose-Einstein condensate act as the two modes of the interferometer.  
The atoms in the two states are trapped in the
wells of a deep one-dimensional optical lattice (deep enough that
there is no tunneling coupling).
Absorption imaging at high spatial
resolution allows the direct measurement of the populations in  $|a\rangle$ and  $|b\rangle$ in each well. 
The two hyperfine states are detected after each other with a delay of $780\,\mu$s.
Controlled inter particle interactions using a Feshbach resonance allow for the realization of a nonlinear beam splitter in which the state at its output ports is a coherent spin
squeezed state (Fig. 1b).
Coherent spin squeezed states are a special kind of many-particle entangled states for which the minimal interferometric phase error is $\Delta \varphi =\xi_{S}/\sqrt{N}$ with the coherent spin squeezing factor \mbox{$\xi_{S}^{2}= N \,(\Delta J_{z})^{2}/(\langle J_{x}\rangle^{2}+\langle J_{y}\rangle^{2})$}.~\cite{Sorensen:2001ab, Wineland:1994aa} 
In the symmetric two mode case the total spin is $J=N/2$ for a system consisting of $N$ quanta.
For large $N$ and small population difference,
$n\ll N$, the orthogonal spin components are related to the population
difference, $n$, and relative phase, $\varphi$, between the two modes as follows: $J_{z} = n$,  $J_{x} \approx N/2 \cos(\varphi)$ and $J_{y} \approx N/2 \sin(\varphi)$.
Here we choose a coordinate
system such that $\langle J_{y}\rangle=0$ when measuring the mean spin length $\langle J_{x}\rangle^{2}+\langle J_{y}\rangle^{2}$.
The spin representation allows the visualization of quantum states on a generalized Bloch sphere as shown in figure 1c.

\begin{figure*}[t]
\begin{center}
\includegraphics{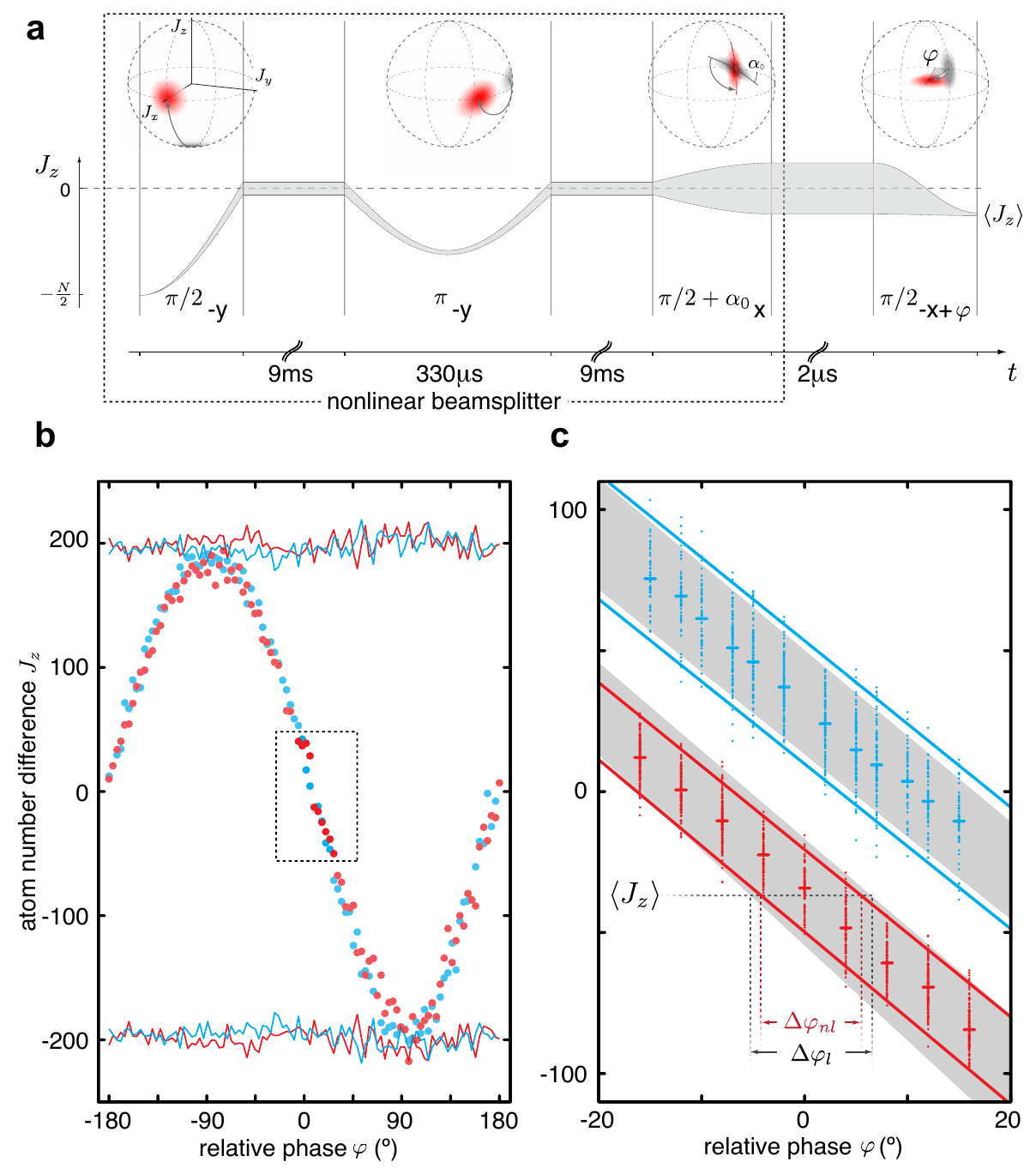}
\caption{Direct experimental demonstration of precision beyond the
standard quantum limit. {\bf a,} The nonlinear Ramsey interferometric sequence in the Bloch sphere representation.
The states before and after the coupling pulses are represented by gray and red shading respectively.
Rotation angles of the different pulses and their rotation axes (subscripts) are as indicated. 
The lower curve represents the temporal evolution of $J_{z}$, its width indicating the
corresponding variance. The nonlinear beam splitter realized with the first three pulses produce a phase squeezed state. 
The last pulse with phase $\varphi$ mixes the $|a\rangle$ and $|b\rangle$ modes before readout of the population imbalance. 
 {\bf b,} A Ramsey fringe observed when scanning the phase $\varphi$ over a full period. 
 The blue data correspond to linear interferometry (the first three pulses are substituted by a single $\pi/2$ pulse) and the red data correspond to nonlinear
interferometry. 
 Technical imperfections cause the decrease in visibility from  $0.98 \pm0.02 $ for the linear to $0.92\pm0.02 $ for the nonlinear interferometer. 
 The solid lines show the total atom number $\pm N/2$ measured for each phase setting as a reference.
 {\bf c,} In order to determine the performance of the nonlinear interferometer we repeat the population difference measurement for a given phase $120$ times (red), concentrating on
the region shown in the dotted box in {\bf b}. 
The solid lines are fits through the lower and upper ends of the two s.d. error bars.
The gray shaded areas show the corresponding uncertainty regions for an ideal linear interferometer with full visibility.
For the nonlinear interferometer, we find a phase error, $\Delta \varphi_{nl}$,
$15\%$ smaller than that, $\Delta \varphi_{l}$, for the ideal linear scheme.
This is remarkable
because for classical linear interferometry (blue), technical noise causes
phase errors $24\%$ larger than those expected for an ideal measurement. For clarity the two measurements are vertically displaced.}
\label{fig.interferometry}
\end{center}
\end{figure*}

For the realization of the nonlinear beam splitter we follow the one-axis-twisting scheme proposed in ref.~\cite{Kitagawa:1993aa}. 
The detailed interferometric sequence is shown in Fig. 2a. 
A fast $\pi/2$ pulse produces a coherent spin state with $\langle J_{z} \rangle = 0$, and subsequent evolution under
the influence of interactions causes a shearing effect transforming
the circular uncertainty region into an elliptical one as detailed in
Fig. 3a.
The resulting quantum state is a coherent spin squeezed state
where the squeezed direction forms at an angle $\alpha_{0}$ relative to the $J_{z}$ axis. 
The final step to realize the nonlinear beam splitter is the rotation of the uncertainty ellipse around its center by $\alpha=\alpha_{0}+\pi/2$ to prepare a phase squeezed state. 
The quality of the output state critically depends on the choice of the axis about which this rotation is made. 
Therefore, we add a spin echo pulse in the evolution period to correct for technical dephasing. 
After the beam splitting process, we allow for $\tau=2\mu s$ of phase accumulation time and then recouple
the states $|a\rangle$ and $|b\rangle$ using a $\pi/2$ pulse with controlled phase $\varphi$ before readout of the population imbalance. \\

The Ramsey fringe resulting from this nonlinear interferometric sequence is shown in Fig. 2b (red), and from it we deduce a visibility of $\mathcal{V} = 0.92 \pm0.02$. 
For a direct comparison with the linear interferometer, we replace the nonlinear beam splitter with a standard linear one -- realized using a single $\pi/2$ pulse -- and keep the total number of atoms at the output constant. We obtain a fringe with a visibility of $\mathcal{V} = 0.98 \pm0.02$ (blue). 
The solid lines above and below the two fringes in Fig. 2b show the total spin length, $\pm N/2$, for reference. 
In order to determine the phase error, we analysed the region around the zero
crossing in more detail, as shown in Fig. 2c. 
We performed $120$ experimental runs per phase setting and measured $\Delta J_{z}^{2}$ for a linear interferometer (blue) and nonlinear interferometer (red). 
It is important to note that none of the data shown in Fig. 2c has
been corrected for noise.
The gray shaded areas indicate the two standard deviation bounds for an ideal classical measurement, i.e. assuming no excess noise due to the measurement process. 
The solid lines are linear fits to the upper and lower ends of the two s.d. error bars of the measured data. 
The difference between the slopes of these lines and
the ideal measurement is caused by the slightly decreased visibility.
The performance of this interferometer in
phase estimation is $15\%$ superior to that in the ideal classical case,
as shown by the dashed lines indicating the respective phase errors in
Fig. 2c. This is remarkable because the phase error that we measure
for the linear interferometer is $24\%$ higher than that expected in an
ideal apparatus.
This excess noise is due to readout noise in the detection scheme~\cite{Esteve:2006aa,Esteve:2008aa} and experimental imperfections in the coupling pulses (see supplementary information).

The experimental sequence starts with a Bose-Einstein condensate of $2300$ $^{87}$Rubidium atoms in the $|F,m_{F}\rangle = |1,-1\rangle$ hyperfine state in an optical dipole trap. 
By splitting the trap into six using a one-dimensional optical lattice potential~\cite{Esteve:2008aa} we are able perform six independent experiments in parallel (Fig. 1c). 
This results in increased statistics for a given measurement time. 
The single traps are almost spherical with dipole frequencies of $\omega_{trap}=2\pi \ast 425\,$Hz. 
An adiabatic passage is used to sweep the atoms to the state $|a\rangle=|1,1\rangle$. 
By two-photon combined microwave and radio frequency pulses we couple the $|a\rangle$  and $|b\rangle=|2,-1\rangle$ states with a Rabi frequency $\Omega$.
The single photon detuning to the $|2,0\rangle$ state is $\delta=-200\,$kHz. 
The chosen states have a narrow Feshbach resonance at $B=9.10\,$G which allows us to control the interspecies s-wave scattering length $a_{ab}$ and therefore to access the miscible regime~\cite{Widera:2004aa, Erhard:2004aa, Kaufman:2009aa,Sinatra:2000aa}. 

For our trap geometries and interaction regime the single spatial mode approximation has been shown to be well justified~\cite{Li:2009aa}. 
The dynamics is described by the Josephson Hamiltonian $H/\hbar=\chi J_{z}^{2}+\Omega J_{\gamma} + \Delta \omega_{0} J_{z}$ where $J_{\gamma} = \cos(\gamma) J_{x} + \sin(\gamma) J_{y}$ is the angular momentum direction in which the coupling pulses are applied and  $\chi \propto a_{aa}+a_{bb}-2 a_{ab}$ is the effective nonlinearity arising from intra- and interspecies interaction.~\cite{Steel:1998aa} 
The ratio of the background scattering lengths $a_{aa}:a_{ab}:a_{bb} = 100:97.7:95$ results in a very small effective nonlinear interaction, but using the Feshbach resonance decreases $a_{ab}$ such that $\chi = 2 \pi \ast 0.063\,$Hz at $B=9.13\,$G. 
The Rabi frequency $\Omega$ can be switched rapidly between  $\Omega = 0\,$Hz and $\Omega = 2 \pi \ast 600\,$Hz allowing for fast diabatic coupling of the states. 
We keep the magnetic field constant throughout the whole interferometric sequence but during the coupling pulses the interaction $\chi J_{z}^{2}$ can be neglected since the system is in the Rabi regime, i.e. $\chi N/\Omega \ll1$. 
Differential energy shifts $\Delta\omega_{0}$ between the $|a\rangle$  and $|b\rangle$ states are mainly due to uncontrolled magnetic field fluctuations. 
We use active magnetic field stabilization to suppress the low frequency components of the magnetic noise to sub milli-gauss levels. 
This is of utmost importance because the performance of the nonlinear beam splitter crucially relies on the final rotation around an axis stable through the center of the noise ellipse.

\begin{figure*}[t]
\begin{center}
\includegraphics[width = 170mm]{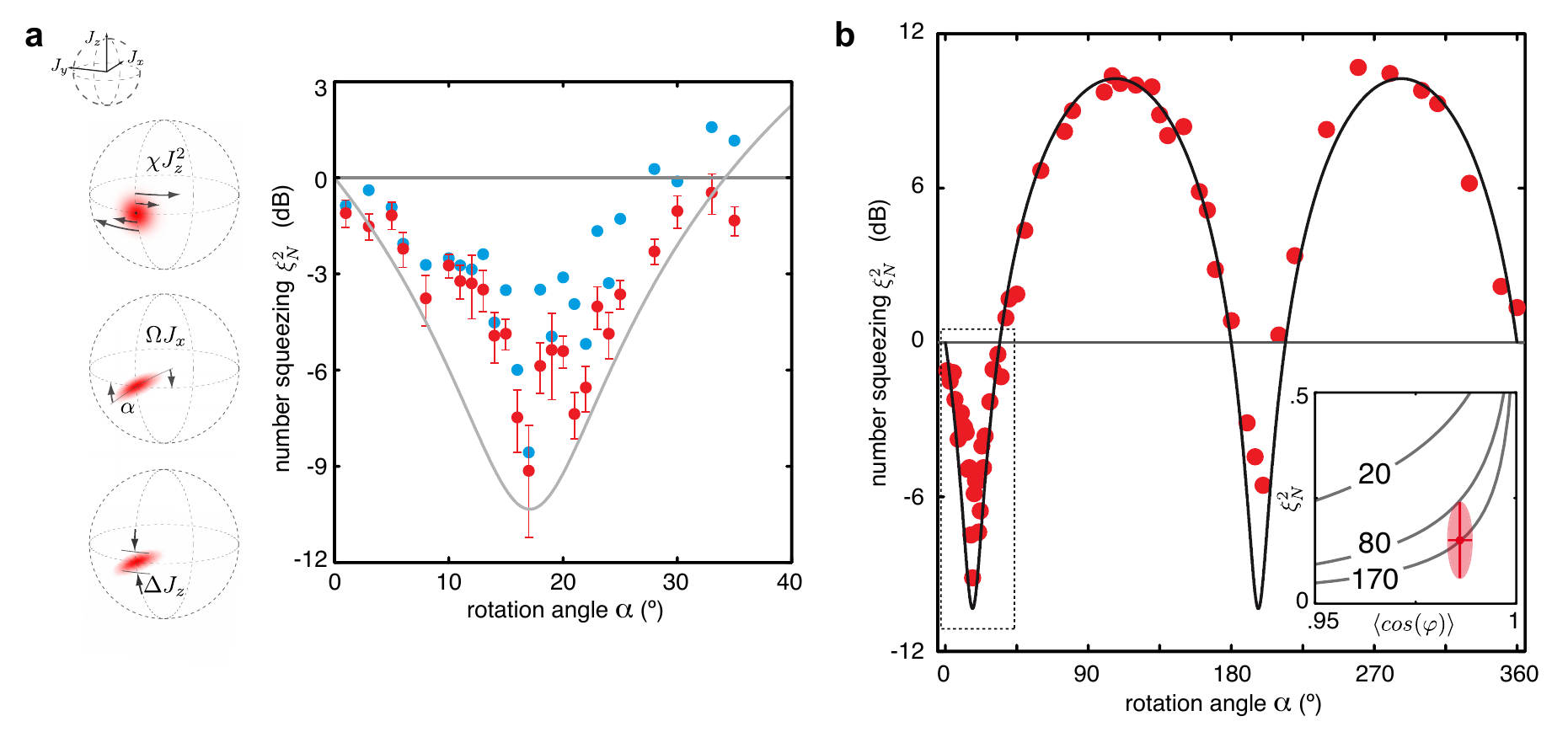}
\caption{Characterization of the quantum state within the nonlinear
interferometer. {\bf a,} One-axis-twisting dynamics of a coherent spin state evolving under a pure $\chi J_{z}^{2}$ Hamiltonian. 
By analogy with the non-interacting
Hamiltonian $H/\hbar = \Delta\omega_{0}J_{z}$, we expect eigenstates of $J_{z}$ to rotate around the $J_{z}$ axis at a rate proportional to $\partial H /\partial J_{z}$. In the interacting case, this gives a
rotation rate proportional to $J_{z}$. Because a coherent spin state with
$\langle J_{z} \rangle < N/2$ can be described as a superposition of several of these
eigenstates, the twisting effect shown in the left-hand column appears.
After a fixed evolution time of $18\,$ms a coupling pulse rotates the quantum state around its center and the fluctuations $\Delta J_{z}$ are detected. 
The graph shows the observed number squeezing factor $\xi_{N}^{2}$ at the output of this nonlinear beam splitter versus rotation angle $\alpha$. 
The blue data have been corrected
for photon shot noise and the red data additionally take the technical noise
into account (Supplementary Information). 
For clarity one s.d. error bars are given for the red data only. 
We observe a best number squeezing factor of $\xi_{N}^{2}=-8.2^{+0.9}_{-1.2}$dB. 
{\bf b,} Noise tomography of the output state of the nonlinear
beam splitter. The dotted box indicates the region detailed in {\bf a}.
The largest fluctuations we measure have a number squeezing factor of $\xi_{N,max}^{2} = +10.3^{+0.3}_{-0.4}\,$dB, resulting in an uncertainty product of the conjugate variances $1.6 \pm0.35$ times larger than expected for a minimal uncertainty state. 
The black line is a fit to the data allowing for one free parameter in the two-mode approximation.  
Because the theory does not include the $15\%$ atom loss, it overestimates the optimal suppression of number fluctuations.
Nevertheless we find good agreement confirming the expected interdependence between number squeezing and purely interaction-driven phase dispersion.
Knowledge of the minimal and maximal spin fluctuations in two orthogonal directions allows for a statement on the many body entanglement present in the system~\cite{Sorensen:2001ab}. 
The inset shows theoretical limits for $\xi_{N}^{2}$  for different minimal non-separable block sizes of the $N$-particle density matrix (gray lines)~\cite{Sorensen:2001aa}. These sizes equal the numbers of entangled atoms.
The red data point is the result of the noise tomography and indicates
entanglement of more than $80$ atoms with a three s.d. confidence level.}
\label{fig.twisting}
\end{center}
\end{figure*}

In addition to the interferometry experiment we characterize the quantum state produced by the nonlinear beam splitter by noise tomography~\cite{Fernholz:2008aa}.  A rotation by $\alpha$ around the center of the quantum state maps spin directions orthogonal to the mean spin orientation to the $J_{z}$ readout axis such that their fluctuations can be measured.
Figure 3a shows a close-up of the results for small angles.
We find minimal fluctuations for $\alpha_{0} = 16 \degree $; the corresponding number
squeezing factor is $\xi_{N}^{2} = 2 \Delta J_{z}^{2}/J =-6.9 ^{+0.8} _{-0.9}$dB, where we average the results for $\alpha = 16 \degree $ and $\alpha = 17 \degree $  to increase the statistics to 634 measurements. 
Here photon shot noise due to the imaging process is removed (blue data points)~\cite{Esteve:2006aa, Esteve:2008aa}. 
The red data in Fig. 3a show the result of also subtracting technical noise due to coupling-pulse imperfections
and magnetic field fluctuations (Supplementary Information).
Using this correction, we infer a number squeezing factor of $\xi_{N}^{2} =  -8.2 ^{+0.9} _{-1.2}\,$dB, which is close to the atom loss limited theoretical optimum for our internal system of $\approx 10\,$dB~\cite{Li:2009aa}. 

Maximum fluctuations of $\xi_{N,max}^{2} = +10.3^{+0.3}_{-0.4}\,$dB are observed for
measurement along the axis orthogonal to the squeezed direction and
limit the coherence, and visibility, of the nonlinear beam splitter
output to $\mathcal{V}=\langle cos(\varphi) \rangle=e^{-\xi_{N, max}^{2}/2 N} = 0.986 \pm0.001$, assuming a gaussian distribution of the phase fluctuations and the validity of the two-mode approximation (see supplementary information) \cite{Ferrini:2009aa}.
The measured Heisenberg uncertainty product $4 \, \Delta J_{z}^{2} \, \Delta J_{y}^{2}/\langle J_{x} \rangle^{2}  $ of the conjugate variances is $1.65 \pm0.35$, which is only slightly larger
than the one expected for a minimal-uncertainty state.
A fit to the data assuming  evolution under the one axis twisting Hamiltonian~\cite{Kitagawa:1993aa} with  the nonlinearity $\chi$ as a free parameter shows very good agreement with theory (Fig. 3b) for $\chi = 2 \pi \ast 0.063\,$Hz.
The discrepancy between fit and data in the number squeezed region is mainly due to a loss of $15\%$ of the atoms during the nonlinear evolution ~\cite{Li:2008aa, Li:2009aa}. 
To put the best number squeezing factor, $\xi_{N}^{2} =  -8.2\,$dB in context with the interferometer experiment, it is important to note that there the nonlinear beam splitter performance was $\xi_{N}^{2} =  -4.3\,$dB  or $\xi_{N}^{2} = -2.1\,$dB, where the first value was inferred from the population-
difference fluctuations at the interferometer output with subtraction
of known technical noise and the second value was inferred
without subtraction of known technical noise. The degradation of the
performance can be explained by the drift of the magnetic fields
(approximately $5\,$mG per day) during the measurement period ($24$ h here but only $3$ h in the measurement of the best number
squeezing).

With knowledge of the fluctuation in two conjugate spin directions,
it is possible to make statements about entanglement in the
spin system~\cite{Sorensen:2001aa, Sorensen:2001ab, Toth:2007aa}. 
For distinguishable particles, entanglement is
defined as the non-separability of the overall density matrix. The
problem of indistinguishability of the atoms in a Bose-Einstein condensate
can in principle be overcome by a local operation whereby
the particles are localized to distinguishable positions in space without
affecting the spin properties of the collective system~\cite{Sorensen:2001aa}. Since local measurements can not generate entanglement it must have been present beforehand.  
The values of squeezed and antisqueezed fluctuations
in two orthogonal spin directions imply a lower bound for
the block size of the largest non-separable part of the density matrix~\cite{Sorensen:2001aa}. 
The measured number squeezing $\xi_{N}^{2}$ and coherence $\langle \cos(\varphi) \rangle$ imply entanglement of $170$ atoms (red data point in inset of Fig. 3b), and we can exclude entanglement of fewer than $80$ atoms with a $3$-s.d. confidence level.

We have directly demonstrated the feasibility of nonlinear atom interferometry beyond the standard quantum limit using a macroscopic
ensemble of atoms. 
Precise characterization of the output state of the nonlinear beam splitter implies coherent spin squeezing of $\xi_{S}^{2} = \xi_{N}^{2}/\langle cos(\varphi) \rangle^{2} =  -8.2$dB.  In principle (that is, if
there is no excess technical noise), this allows for a $61\%$ increase in
phase sensitivity over classical linear interferometry. This is a significant
step towards useful spin squeezing in atom interferometry~\cite{Fernholz:2008aa, Appel:2009aa, Schleier-Smith:2010aa, Esteve:2008aa}. 
The extension of many particle
atom interferometry to the nonlinear regime is thus an
advance towards applied quantum atom optics using coherent interactions
between atoms.
We note that the group of P. Treutlein has independently realized internal-state spin squeezing on an atom chip
through controlled interactions using state-dependent potentials~\cite{Riedel:2010aa}.
 
\newpage

\appendix
\section{Supplementary Information}
\subsubsection{Calculation of the atom number difference fluctuations}  \label{sec.deltan}
We deduce the population of each hyperfine state by state selective high intensity absorption imaging (see next section).
For each well, the raw data consists of a set of atom numbers $N_a$ and $N_b$ in the two hyperfine states $|a\rangle$ and $|b\rangle$ obtained by repeating the experiment 60 times. The data shown in figures 3 and 4 of the main paper are average numbers extracted from several wells (with total atom number between $200$ and $450$) and over several datasets measured at different days. The indicated error bars and error values are obtained as the statistical standard error of the mean obtained from the described averaging. A Grubb outlier detection algorithm [F.E. Grubbs, Technometrics {\bf 11} 1 (1969)] is used to filter the extracted atom number difference $(N_a - N_b)/2$. It detects typically zero but maximally $1$ to $2$ points out of $60$ (at a $5\%$ significance level). Due to possible slow drifts of the magnetic field (on the hour timescale) we correct each dataset and remove a linear slope. Statistical simulations were performed to test this procedure and biasing was found to be negligible.\\
For each dataset, we define $p = \langle N_a/N_{tot}\rangle$ the probability for an atom to be in state $|a\rangle$ where $N_{tot} = N_a + N_b$ is the total atom number. 
If $p \neq 1/2$ the atom number difference $n$ depends on the total atom number as $n = (p-1/2) N_{tot}$. Therefore fluctuations in the total atom number between different experimental runs (which are on the order of $\sqrt{N_{tot}}$), contribute to the measured variance. We compute 
\begin{equation}
\Delta n_{raw}^2 = \langle \left[(N_a - N_b)/2 - (p-1/2)N_{tot}\right]^2 \rangle \nonumber
\end{equation}
in order to avoid taking into account these fluctuations. Since $p$ is typically close to $1/2$ this correction has only a small effect.
Additional noise $\delta N_{a (b),{\rm psn}}^{2}$  in the atom number $N_{a}$  ($N_{b}$) per state due to photon shot-noise from the detection process contributes to the variance $\Delta n_{raw}^{2}$.
We deduce this extra variance as the sum over all CCD pixels in the integration area where the contribution per pixel is inferred from the light intensity on the absorption and on the reference picture.
A measured CCD camera noise calibration curve which relates the variance per pixel to the mean counts permits to calculate the additional atomic variances $\delta N_{a (b),{\rm psn}}^{2}$  for each experimental realization.
We subtract this contribution 
\begin{equation}
\Delta n_{\rm psn}^{2} = [1/4 + (p-1/2)^2] \langle \delta N_{a,{\rm psn}}^2+\delta N_{b,{\rm psn}}^2 \rangle \nonumber
\end{equation}
and obtain the corrected number fluctuations
\begin{equation}
\Delta n^2  = \Delta n_{raw}^2 -\Delta n_{\rm psn}^{2} \nonumber
\end{equation}
The typical magnitude of $\Delta n_{\rm psn}$ corresponds to fluctuations of $6.8$ atoms. For the best measured number squeezing $\xi_{N}^{2}=-6.9$dB, the total fluctuations are $\Delta n_{raw} \approx 8.1$ atoms, while the atomic shot noise limit is $9.8$ atoms. Even though the photon shot noise gives a large contribution to the total fluctuations its subtraction is very accurate. The uncertainty in $\Delta n_{raw}^2$ is $4\%$ (two s.d. error) and the main contribution is due to statistical errors in the CCD camera noise calibration curve.\\
Finally the number squeezing  factor $\xi_N^2$ is calculated by normalizing $\Delta n^2$
by the expected value for a coherent spin state with population ratio $p$ between the states $|a\rangle$ and $|b\rangle$
\begin{equation}
\xi_N^2 =  \frac{\Delta n^2}{p(1-p) \langle N_{tot}\rangle} \nonumber
\end{equation}
We emphasize that no post processing (i.e. no photon shot noise, drift removal or total atom number fluctuation correction) is possible in the interferometric measurement shown in figure 2c of the main paper. The plotted data  in figure 2c is directly $\Delta n = (N_a - N_b)/2$ after the Grubbs test.

\begin{figure}
\begin{center}
\vspace{10mm}
\includegraphics[width = 8.5cm]{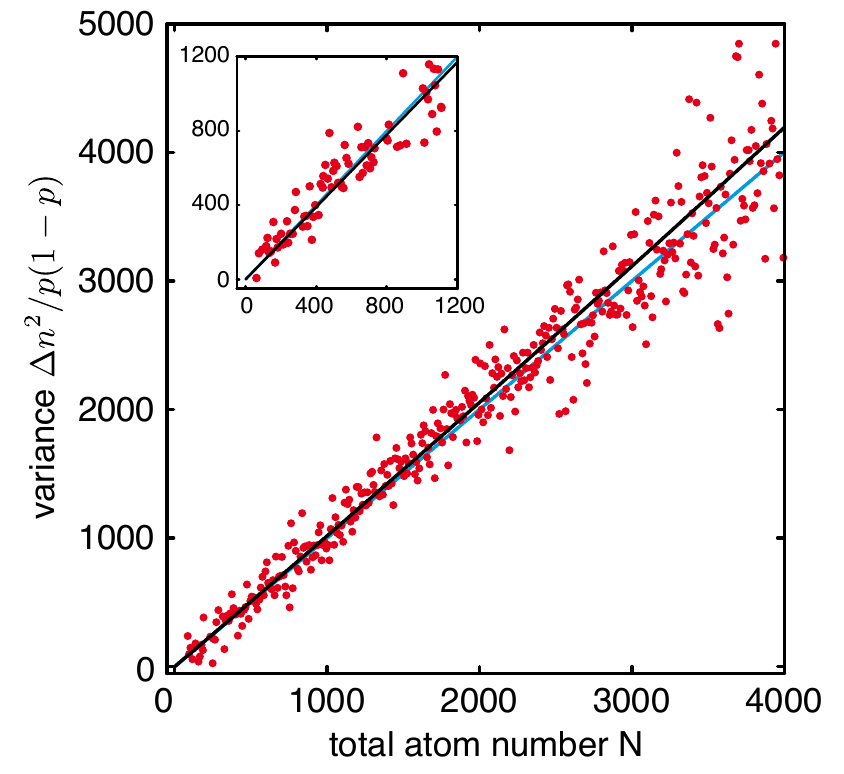}
\caption{Correct calibration of the atom number detection is crucial for the results presented in this paper. We performed an independent check of the calibration by measuring the atom number difference fluctuation of a coherent spin state versus the total atom number. The expected linear slope for a coherent spin state is unity (blue line). The resulting fluctuations analyzing all different wells of the optical lattice are plotted in the inset. A linear fit (black line) reveals a slope of $0.98\pm0.06$. In order to increase the statistics and to expand the total atom number range we sum over a different number of wells and calculate the fluctuations from the binned atom numbers. We repeat this procedure for all possible combinations of the binning. The high statistics allows for a second order fit (black line) and we find $1.01 \pm0.03$ as the linear coefficient and a very small quadratic contribution of $2\ast10^{-5}$. All uncertainties given are two s.d. errors.}
\label{fig.calibration}
\end{center}
\end{figure}

\subsubsection{Atom number measurement and calibration}
We detect the atoms by high intensity absorption imaging. The imaging system has a spatial resolution of $\approx 1\mu$m allowing us to unambiguously resolve the individual lattice sites spaced by $5.7\mu$m. For precise calibration of the total atom number we follow the procedure described in~[J. Est\`eve \emph{et al.}, Nature {\bf 455} 1216 (2008)] and in~[G. Reinaudi \emph{et al.}, Opt. Lett. {\bf 32} 3143 (2007)]. \\
The detection sequence starts with ramping down the magnetic offset field from $9.13$G, its value during the experimental sequence, to a value close to zero within $3$ms. Afterwards the optical dipole trap levitating the atoms against gravity is switched off  to expand the atomic clouds. In order to avoid overlapping of clouds from different wells during this expansion the optical lattice is kept on.  $570\mu$s after dipole trap switch off the $|2,-1\rangle$ atoms are imaged with a 10$\mu$s to $25\mu$s long imaging pulse. The imaging pulse intensity is $I\approx 10 I_{sat}$ to optimize the signal to noise ratio ($I_{sat}$ is the saturation intensity). The detected atoms are removed from the field of view of the camera by an additional resonant laser pulse and the $|1,1\rangle$ atoms are optically pumped to the $F=2$ groundstate hyperfine manifold.  $780\mu$s after the first imaging pulse a second pulse with the same parameters is used to detect the second species. \\
The result of an independent test of the imaging calibration is shown in figure 1. 
We prepare a coherent spin state by a fast $\pi/2$ pulse and measure the fluctuations of the population difference $\Delta n^{2}/p(1-p)$ versus the total atom number $N$. 
For a coherent spin state we expect $\Delta n^{2}/p(1-p) = N$. In order to vary the total atom number and test the scaling of the fluctuations we use two approaches: We analyse the different wells of the optical lattice individually and we also vary the total atom number in the six well system. This covers the range from zero to $\approx 1000$ atoms in the different wells. The result of this analysis is shown in the inset in figure 1. The fitted linear slope is $0.98\pm0.06$ (two s.d. error). In a second step we analyze the data after binning a different number of wells (see next section). The same fluctuation calculation is done for every possible permutation of the binning. This method results in high statistics and widens the range of atom numbers for which we test our imaging. A quadratic fit yields a linear coefficient of $1.01 \pm0.03$ (two s.d. error) while the quadratic one is $2\ast10^{-5}$. This small nonlinearity is caused by errors in the preparation pulse or by excess noise in the imaging procedure. As a result of this imaging test we find an upper limit of $4\%$ for a possible systematic error due to wrong calibrations in the detection procedure with $95\%$ confidence level.

\begin{figure}[b]
\begin{center}
\vspace{10mm}
\includegraphics[width = 8.5cm]{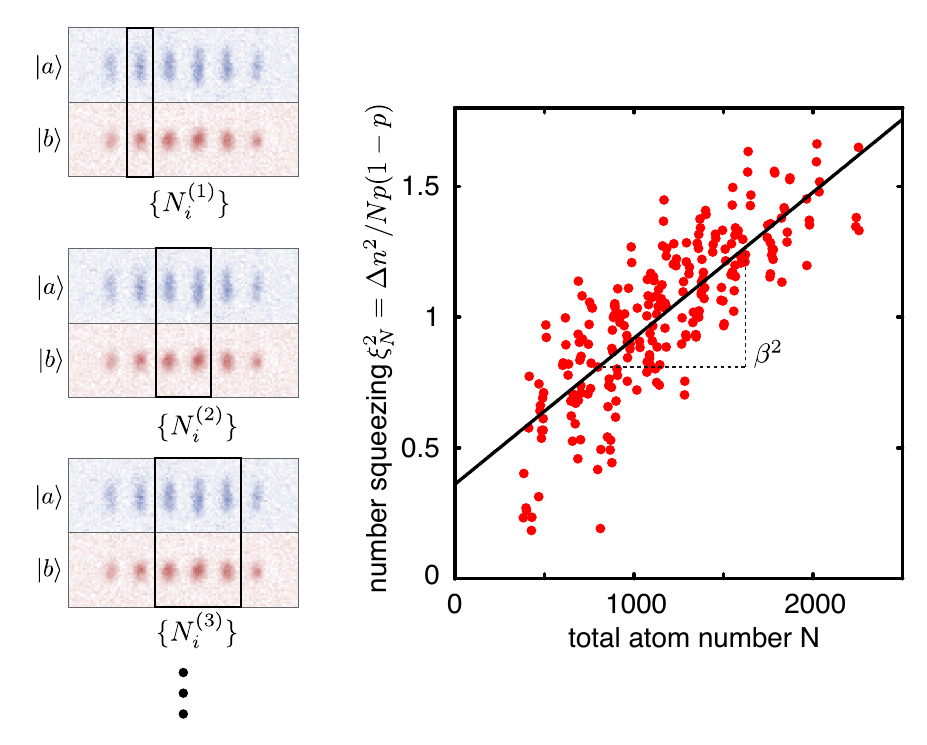}
\caption{The left panel illustrates the binning procedure to obtain different total atom numbers from a single dataset. We sum the region in the pictures of the $|a\rangle$ ($|b\rangle$) atoms corresponding to $k$ wells and obtain the atom number $N_{a}^{(k)}$ ($N_{b}^{(k)}$). For each binning the total atom number is given by $N^{(k)}=N_{a}^{(k)} + N_{b}^{(k)}$. We repeat the procedure for every possible permutation and calculate the normalized fluctuations for each total atom number $\{N_{i}^{(k)}\}$. Plotting the resulting squeezing $\xi_{N}^{2}$ versus the total atom number typically reveals a linear correlation.  This correlation is due to technical fluctuations affecting all wells in the same way, since the condensates in different wells are separated. The fitted linear slope $\beta^{2}$ is used to subtract the resulting excess noise from the measured number squeezing in a single well.}
\label{fig.technoise}
\end{center}
\end{figure}

\subsubsection{Technical noise sources} \label{sec.technoise}
Uncontrolled fluctuation of the magnetic field during the experimental sequence or coupling pulse errors due to frequency or timing noise results in a contribution to $\Delta n^{2}$ scaling as $\Delta n_{tech}^{2} \propto \beta^{2} N^{2}$. $\beta^{2}$ measures the integrated magnitude of the technical noise. The unfavorable scaling with $N$ is one of the main challenges to achieve squeezing systems with larger atom number. \\
We perform six independent experiments in parallel, well separated by a high potential barrier. Crosstalk between the condensates within the experimental timescale of $20$ms is excluded since the tunneling timescale between the wells is in the order of seconds. Since all wells are exposed to the same noise, correlation between them can be used to extract the noise $\beta^{2}$. For each dataset taken we bin different numbers of wells and analyse the resulting number fluctuations between the two internal states  for all possible permutations of the binning (see figure 2). We plot the calculated number squeezing $\xi_{N}^{2}(N)$ versus the total atom number $N$ contained in the binned area of the picture. Technical noise results in a non zero slope $\beta^{2} $ which we extract by a linear fit to the data as shown in figure 2. The technical noise contribution to the measured number squeezing for each lattice site with total atom number $N$ is $\xi^{2}_{tech} =\beta^{2} N$. In order to measure the number difference fluctuations of the quantum state we subtract this technical contribution.

\begin{figure}[b]
\begin{center}
\vspace{10mm}
\includegraphics[width = 8.5cm]{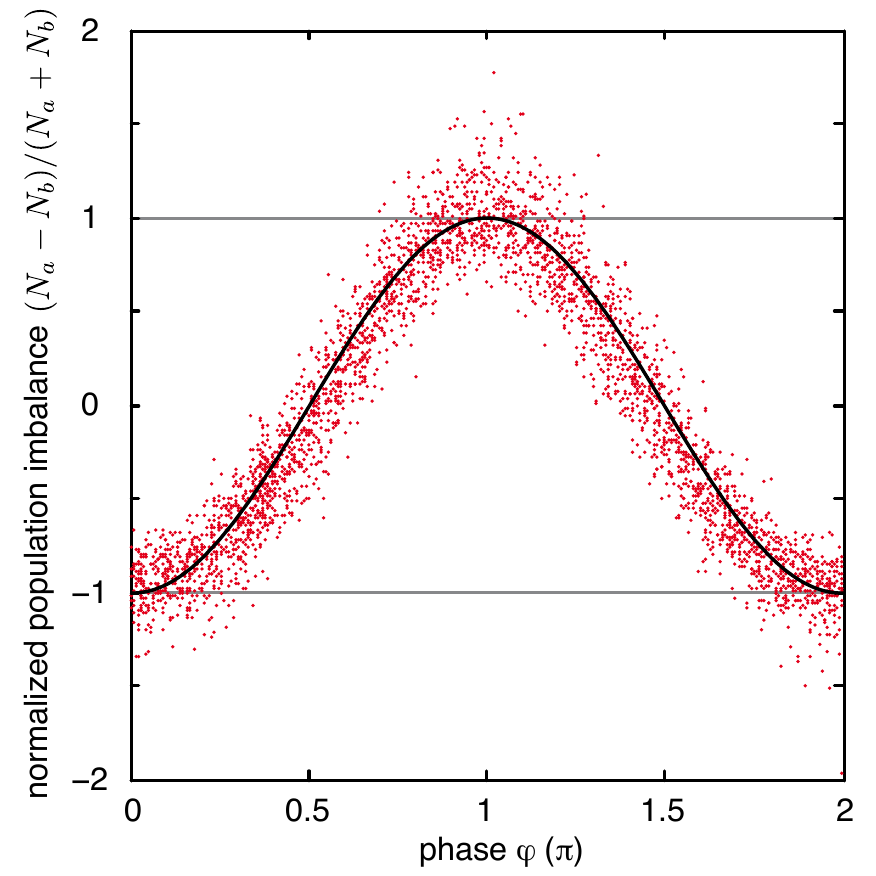}
\caption{A Ramsey fringe measured at the output of the nonlinear beam splitter. A sinusoidal fit reveals a visibility of $100\%$ with a statistical uncertainty of $2\%$.}
\label{fig.technoise}
\end{center}
\end{figure}

\subsubsection{Validity of the two-mode approximation}
In the symmetric subspace the total spin length $J=N/2$ is given by the total number of atoms $N$. Here the mean spin length $\langle J_{x} \rangle=N/2 \langle \cos(\varphi) \rangle$ can be calculated from the total number of atoms and from the fluctuations in the most uncertain orthogonal direction $\Delta J_{y}$. In order to check the validity of the two mode approximation, we compare the visibility $\mathcal{V} = \langle \cos(\varphi) \rangle$ obtained from a Ramsey fringe measurement $\mathcal{V}_{R}$ to the visibility calculated under the two mode assumption $\mathcal{V}_{2}$. Figure 3 shows a Ramsey fringe measured at the output of the nonlinear beam splitter. A sinusoidal fit reveals $\mathcal{V}_{R} = 1.00 \pm 0.02$ which is consistent with the visibility deduced from the noise tomography measurement,  justifying the two-mode approximation. Since the noise measurements can be done with higher reliability and accuracy we use $\mathcal{V}_{2} = e^{-\xi_{N, max}^{2}/2 N}  = 0.986 \pm0.001$ to calculate the visibility in the main paper.

\begin{acknowledgments}
We thank Jens-Philipp Ronzheimer for technical assistance throughout the realization of this project and acknowledge fruitful discussions with Yun Li and Alice Sinatra. We gratefully acknowledge support from the DFG, GIF and the EC FET-Open MIDAS  project. C.G. acknowledges support from the Landesgraduiertenf\"orderung Baden-W\"urttemberg. \\
Correspondence should be addressed to M.K.O.~(email: quantum.metrology@matterwave.de).
\end{acknowledgments}

\bibliography{interferometrybib.bib}

\end{document}